\documentclass[prd,showpacs,showkeys,eqsecnum,nofootinbib,preprint,superscriptaddress]{revtex4-1}
\usepackage[utf8]{inputenc}
\usepackage{epsfig}
\usepackage{amsmath}
\usepackage{amsfonts}
\usepackage{amssymb}
\usepackage{color}
\usepackage{graphicx}
\usepackage{url,hyperref}

\begin{document}
\title{{Naturalness in Type II seesaw model and implications for physical scalars}}
\author{M.~Chabab}
\email{mchabab@uca.ma}
\affiliation{Laboratoire de Physique des Hautes Energies et Astrophysique \\
D\'epartement de Physiques, FSSM, Universit\'e Cadi Ayyad, Marrakech, Morocco}
\author{M.~C. Peyran\`ere}
\email{michel.capdequi-peyranere@univ-montp2.fr}
\affiliation{Universit\'e de Montpellier, Laboratoire Univers \& Particules de 
Montpellier, F-34095 Montpellier, France}
\affiliation{CNRS/IN2P3, Laboratoire Univers \& Particules de 
Montpellier UMR 5299, F-34095 Montpellier, France}
\author{L.~Rahili}
\email{rahililarbi@gmail.com}
\affiliation{Laboratoire de Physique des Hautes Energies et Astrophysique \\
D\'epartement de Physiques, FSSM, Universit\'e Cadi Ayyad, Marrakech, Morocco}

\begin{abstract}
\noindent 
In this paper we consider a minimal extension to the standard model by a scalar triplet field with hypercharge $Y=2$.
This model relies on the seesaw mechanism which provides a consistent explication of neutrino mass generation. We show 
from naturalness considerations that the Veltman condition is modified by virtue of the additional scalar charged states 
and that quadratic divergencies at one loop can be driven to zero within the allowed space parameter of the model, the latter
 is severely constrained by unitarity, boundedness from below and is consistent with the di-photon Higgs decay data of LHC. 
Furthermore, we analyse the naturalness condition effects to the masses of heavy Higgs bosons $H^0$, $A^0$, 
$H^\pm$ and $H^{\pm\pm}$, providing a drastic reduction of the ranges of variation of $m_{H^{\pm}}$ and $m_{H^{\pm\pm}}$ 
with an upper bounds at $288$ and $351$~GeV respectively, while predicting an almost degeneracy for the other neutral Higgs
 bosons $H^0$, $A^0$ at about $207$~GeV.
\end{abstract}
\maketitle	
\section{Introduction}
\label{sec:introd}
After the LHC's Run 1 and beginning of Run 2, we are now more confident that the observed $125$ GeV scalar boson  is the long sought Higgs boson of the Standard Model (SM) \cite{atlas12, cms12}. However, although its brilliant success in
 describing particle physics, still many pressing questions are awaiting convincing solutions that cannot be answered within SM. The hierarchy problem and the neutrinos oscillations are the most illustrative ones. In this context, many theoretical frameworks have been proposed and the most popular one is Supersymmetry.

The search for Supersymmetry at Run I of LHC gave a negative result. Therefore the original motivation of Susy to solve hierarchy problem by suppressing quadratic divergencies (QD) is questionnable. In this case,  it is legitimate to propose other perspective to interpret and control the QD. It is known that one has to call upon new physics to deal with such problem. More specifically, the new degrees of freedom in a particular model conspire with those of the Standard Model to modify the Veltman Condition and to soften the divergencies  \cite{drozd,masina, kundu13, biswas14}. 

In this paper, we aim to investigate the naturalness problem in the context of Type II Seesaw model, dubbed HTM,  with emphasis on its effect of the HTM parameter space . More precisely, we will study how to soften the divergencies and how to gain some insight on the allowed masses of the heavy scalars in the Higgs sector.  A more recent work of Kundu et al.\cite{Kundu14} has partially discussed this issue. However, unlike the analysis in \cite{Kundu14}, our study use the most general renormalisable Higgs potential of HTM \cite{aa11} and is essentially based on dimensional regularisation approach which complies with unitarity and Lorentz invariance \cite{aa11}.  More importantly, the phenomenological analysis takes into account the full set of theoretical constraints, including unitarity \cite{aa11} and the consistent conditions of boundedness from below  \cite{aa11, spanish15}. 

This work is organised as follows. In section $2$, we briefly review the main features of Higgs Triplet Model and present the full set of constraints on the parameters of the Higgs potential. Section $3$ is devoted to the derivation of the modified Veltman condition (mVC) in HTM. The analysis and discussion of the results are performed in section $4$, with emphasis on the effects of mVC on the heavy Higgs bosons, particularly on charged Higgs. Conclusion with summary of our results will be drawn in section $5$.
%
\section{Seesaw Type II Model: brief review}
\label{sec:reviewmodel}

Type II seesaw mechanism can be implemented in the Standard Model via a scalar field $\Delta$ transforming as a triplet 
under the $SU(2)_L$ gauge group with hypercharge $Y_\Delta=2$. In this case  
the most general $SU(2)_{L}\times U(1)_{Y}$ gauge invariant Lagrangian of the {\rm HTM} scalar sector  
is given by  \cite{aa11,Perez08}:

\begin{eqnarray}
\mathcal{L} &=&
(D_\mu{H})^\dagger(D^\mu{H})+Tr(D_\mu{\Delta})^\dagger(D^\mu{\Delta}) -V(H, \Delta) + \mathcal{L}_{\rm Yukawa}
\label{eq:DTHM}
\end{eqnarray}
\noindent
The covariant derivatives are defined by,  
\begin{eqnarray}
D_\mu{H} &=& \partial_\mu{H}+igT^a{W}^a_\mu{H}+i\frac{g'}{2}B_\mu{H} \label{eq:covd1}\\
D_\mu{\Delta} &=&
  \partial_\mu{\Delta}+ig[T^a{W}^a_\mu,\Delta]+ig' \frac{Y_\Delta}{2} B_\mu{\Delta} \label{eq:covd2}
\end{eqnarray}
\noindent
where $H$ is the Higgs doublet while (${W}^a_\mu$, $g$), and ($B_\mu$, $g'$) represent the $SU(2)_L$ and $U(1)_Y$ gauge fields and couplings respectively. 
$T^a \equiv \sigma^a/2$, with $\sigma^a$ ($a=1, 2, 3$)  are the Pauli matrices.
The potential $V(H, \Delta)$ reads as,
\begin{eqnarray}
V(H, \Delta) &=& -m_H^2{H^\dagger{H}}+\frac{\lambda}{4}(H^\dagger{H})^2+M_\Delta^2Tr(\Delta^{\dagger}{\Delta})
+[\mu(H^T{i}\sigma^2\Delta^{\dagger}H)+{\rm h.c.}] \nonumber\\
&&+\lambda_1(H^\dagger{H})Tr(\Delta^{\dagger}{\Delta})+\lambda_2(Tr\Delta^{\dagger}{\Delta})^2
+\lambda_3Tr(\Delta^{\dagger}{\Delta})^2 +\lambda_4{H^\dagger\Delta\Delta^{\dagger}H}
\label{eq:Vpot}
\end{eqnarray}
where $Tr$ denotes the trace over $2\times2$ matrices. The Triplet $\Delta$ and doublet Higgs $H$  
are represented by:
\begin{eqnarray}
\Delta &=\left(
\begin{array}{cc}
\delta^+/\sqrt{2} & \delta^{++} \\
\delta^0 & -\delta^+/\sqrt{2}\\
\end{array}
\right) \qquad {\rm and} \qquad H=\left(
                    \begin{array}{c}
                      \phi^+ \\
                      \phi^0 \\
                    \end{array}
                  \right)
                  \label{HDrep}
\end{eqnarray}
with $\delta^0=\frac{1}{\sqrt{2}} (v_t+\xi^0+iZ_2) $ and $\phi^0 = \frac{1}{\sqrt{2}} (v_d+h+iZ_1)$.\\
\noindent
After the spontaneous electroweak symmetry breaking, the Higgs doublet and triplet fields acquire their
vacuum expectation values $v_d$ and $v_t$ respectively, and seven physical
Higgs bosons appear, consisting of: two $CP_{even}$ neutral scalars ($h^0$, $H^0$), one neutral pseudo-scalar $A^0$ and a pair of  simply and doubly charged Higgs bosons $H^\pm$ and $H^{\pm\pm}$. \footnote{For details on mixing angles for $CP_{even}$, $CP_{odd}$ and the charged sectors, dubbed $\alpha$, $\beta$ and $\beta^{'}$ see \cite{aa11} }. The masse of these Higgs bosons are given by \cite{aa11},
\begin{eqnarray}
&& m_{h^0} = \frac{1}{2}(A+C - \sqrt{(A-C)^2 + 4 B^2}) \label{eq:mh0}\\
&& m_{H^0} = \frac{1}{2}(A+C + \sqrt{(A-C)^2 + 4 B^2}) \label{eq:mhh}\\
&& m_{H^{\pm\pm}}^2 = \frac{\sqrt{2}\mu{\upsilon_d^2}-\lambda_4\upsilon_d^2\upsilon_t-2\lambda_3\upsilon_t^3}{2v_t}\label{eq:mHpmpm}\\
&& m_{H^{\pm}}^2 = \frac{(\upsilon_d^2+2\upsilon_t^2)\,[2\sqrt{2}\mu-\lambda_4\upsilon_t]}{4\upsilon_t}\label{eq:mHpm}\\
&& m_{A^0}^2 = \frac{\mu(\upsilon_d^2+4\upsilon_t^2)}{\sqrt{2}\upsilon_t}\label{eq:mA0}
\end{eqnarray}
\noindent
The coefficients $A, B$ and $C$ are the entries of the $CP_{even}$ mass matrix defined by,
\begin{equation}
A=\frac{\lambda}{2}v_d^2,\hspace{0.2cm}
B=v_d(-\sqrt{2}\mu+(\lambda_1+\lambda_4)v_t),\hspace{0.2cm}
C=\frac{\sqrt{2}\mu\,v_d^2+4(\lambda_2+\lambda_3)v_t^3}{2v_t}
\label{ABC:cpeven}
\end{equation}
In the remainder of this paper, we assume the light $CP_{even}$ scalar $h^0$ as the observed Higgs with  mass about $m_{h^0} \simeq 125$~GeV.
\section{Theoretical and experimental constraints}
The HTM Higgs potential parameters are not free but have to obey several constraints originating from theoretical requirements and experimental data. Thus any phenomenological studies are only reliable in the allowed region of HTM parameter space. \\

{\sl \underline{$\rho$ parameter}:} \\
First, recall that the $\rho$ parameter in HTM at the tree level is given by the formula, $\rho \simeq 1 - 2 \frac{v_t^2}{v_d^2}$ which indicates a deviation from unity. Consistency with the current limit on $\rho$ from precision measurements \cite{db12} requires that the limit $|\delta\rho| \leq10^{-3}$ resulting in an upper limit on $v_t$ about $\leq 5$~GeV. \\

{\sl \underline{Masses of Higgs bosons  }:} \\
Many experimental mass limits have been found for the Heavy Higgs bosons. From the LEP direct search results, the lower bounds on $m_{A^0, H^0} >  80-90$ ~GeV for models with more than one doublet in the case of the neutral scalars.  

As to the singly charged Higgs mass we use the LEP II latest bounds, $m_{H^{\pm}} \geq 78$~GeV from direct search results, whereas the indirect limit is slightly higher $m_{H^{\pm}} \geq 125$~GeV \cite{lep}. Furthermore, the present lower bound from  LHC is $m_{H^{\pm}}$  $\leq 666$~GeV, where the excluded mass ranges established by ATLAS \cite{atlas_charged} and CMS \cite{cms_charged} are taken into account.  In the case of the doubly charged Higgs masses, the most recent experimental upper limits  reported by ATLAS and CMS  are respectively $m_{H^{\pm \pm}} \geq 409$~GeV \cite{atlas_dcharged} and $m_{H^{\pm \pm}} \geq 445$~GeV  \cite{cms_dcharged}. These bounds originate from analysis assuming 100\% branching ratio for $H^{\pm\pm}\to l^\pm l^\pm$ decay.  However, note that one can find realistic scenarios where this decay channel is suppressed with respect to  $H^{\pm\pm}\to W^{\pm} W^{\pm (*)}$  \cite{akeroyd08} invalidating partially the LHC limits. For example, In HTM with moderate triplet' VEV,  $v_t \approx 1$~GeV, the analysis of $H^{\pm\pm}\to W^{\pm} W^{\pm *}$ decay channel can easily overpasses the two-sign same lepton channel for $m_{H^{\pm\pm}}$ where the limit decreases up to $90-100$~GeV \cite{goran12}.  \\

As to the theoretical constraints on the parameter space, we should take into account the perturbativity constraints
on the $\lambda_i$ as well as the stability of the electroweak vacuum that ensure that the potential is bounded from below (BFB).
 Let us first recall all the constraints obtained in \cite{aa11}\footnote{Notice here that for BFB,
 the new BFB condition derived by \cite{spanish15} has been used. However, our analysis is almost insensitive to the modified BFB.} 


{\sl \underline{BFB}:}
\begin{eqnarray}
&& \lambda \geq 0 \;\;{\rm \&}\;\; \lambda_2+\lambda_3 \geq 0  \;\;{\rm \&}  \;\;\lambda_2+\frac{\lambda_3}{2} \geq 0 \label{eq:bound1} \\
&& {\rm \&} \;\;\lambda_1+ \sqrt{\lambda(\lambda_2+\lambda_3)} \geq 0 \;\;{\rm \&}\;\; \lambda_1+\lambda_4 \sqrt{\lambda(\lambda_2+\lambda_3)} \geq 0  \label{eq:bound2} \\
&& {\rm \&} \;\; \lambda_3 \sqrt{\lambda} \le |\lambda_4| \sqrt{\lambda_2+\lambda_3} \;\; {\rm or} \;\; 
2 \lambda_1+\lambda_4+\sqrt{(\lambda-\lambda_4^2) (2\frac{\lambda_2}{\lambda_3} + 1)} \geq 0 \label{eq:bound3}
\end{eqnarray}

{\sl \underline{Unitarity}:}
\begin{eqnarray}
&&|\lambda_1 + \lambda_4| \leq 8 \pi  \label{eq:unit1} \\
&&|\lambda_1| \leq 8 \pi \label{eq:unit2} \\
&&|2 \lambda_1 + 3 \lambda_4| \leq 16 \pi \label{eq:unit3} \\
&&|\lambda| \leq  16 \pi \label{eq:unit4} \\
&&|\lambda_2| \leq  4 \pi \label{eq:unit5} \\
&&|\lambda_2 + \lambda_3| \leq  4 \pi \label{eq:unit6} \\
&&|\lambda + 4 \lambda_2 + 8 \lambda_3 \pm \sqrt{(\lambda - 4 \lambda_2 - 8 \lambda_3)^2
+ 16 \lambda_4^2} \;| \leq  32 \pi \label{eq:unit7} \\
&&| 3 \lambda + 16 \lambda_2 + 12 \lambda_3 \pm \sqrt{(3 \lambda - 16 \lambda_2 - 12 \lambda_3)^2
+ 24 (2 \lambda_1 +\lambda_4)^2} \;| \leq  32 \pi \label{eq:unit8} \\
&&|2 \lambda_1 - \lambda_4| \leq 16 \pi \label{eq:unit9} \\
&&|2 \lambda_2 - \lambda_3| \leq  8 \pi, \label{eq:unit10} 
\end{eqnarray}

\noindent
We stress here that all values presented in the plots  and in our subsequent analysis (section $5$) are consistent with all theoretical and experimental bounds described in this section.

\section{The modified Veltman Condition}
\label{sec:veltman-condi}
The method to collect the quadratic divergencies in a framework of dimensional regularisation is due to 
Veltman \cite{Veltman81} and needs a dimensional gymnastics since the space-time dimension to pick up the quadratic 
divergencies depends on the number of loops. But the idea to use the quadratic divergencies to get physical 
insights is much older, and goes back to the pioneering work of Stuckelberg in 1939 \cite{Chakraborty39}. 
In the context of a renormalisable theory (which is our case), it is not clear that the deductions coming 
from such a concept are pertinent. Indeed it provides for relations between 
the masses or/and the couplings constants of the theory, which could give some credit to an underlying and 
hypothetical symmetry.

From a more basic level, one can imagine that the relations found by this method are indicative and could 
provide orders of magnitude, bounds or constraints and we will follow this line in this work. This is more or 
less the point of view of Chakraborty-Kundu \cite{Kundu14} where they apply the cancellation of quadratic 
divergencies to the HTM, the Standard Model extended by a scalar triplet. But we differ 
from this work in the way to pick up the quadratic divergencies. They use the cut-off method to regularise 
the divergent integrals although we will use the Veltman method that we will sketch later on. Both methods 
have advantages and back-draws that we will not discuss here, since they provide for similar relations. 
We just mention that the cut-off method is intuitive but neither Lorentz nor gauge invariant. A contrario, the
dimensional regularisation method is both Lorentz and gauge invariant, although less intuitive. We refer to previous works such as 
\cite{Peyranere1991} and references therein for more details and discussions. 

To check the Veltman conditions one needs to calculate the quadratic divergencies which show up in the tadpoles of the two CP-even 
neutral Higgs of our model, namely $h^0$ and $H^0$. Since vacuum is supposed to be CP-even, the tadpole 
associated with the neutral pseudoscalar field $A^0$ vanishes. Also,  it is noticeable that no QCD 
contribution appears at one loop level, hence only the electro-weak part of the HTM model is concerned in this procedure. Since there is no derivative couplings, 
it is easy to figure out how to calculate each diagram (see the appendix for 
all relevant vertices required for such calculation). Aside the coupling constant, we just need to consider the propagator of the field in the loop, 
where the contribution of each loop can be written in a simple form  in terms of the Passarino-Veltman function
$A_0(m^2) =\frac{i}{16\pi^2} \int \frac{d^nq}{q^2 - m^2}$ \cite{pave}.

In a 4 dimensional space-time, $A_0(m^2)= m^2 (\Delta_4 + 1 - \log m^2)$, up to some pure numerical factor, 
where $\Delta_4$ is the pole term. In a 2 dimensional space-time, $A_0(m^2)$ is no longer mass dimensioned 
and it is a pure U.V. divergent number; this property is fundamental because it guarantees a gauge invariant result. 
This pure number will be forgotten everywhere in this section since irrelevant in the equations, together with
other common factors like the dimensional dependence of the coupling constants.

For the propagator of a scalar particle in HTM Higgs sector, ($h^0, H^0, A^0, G^0, G^{\pm},\eta^Z, \eta^{\pm}$),  
we have $\frac{i}{q^2 - m^2}$ 
leading to a $A_0(m^2)$ contribution. For a vectorial particle ($W^{\pm},Z)$, we have 
$-i (\frac{ T^{\mu \nu}}{q^2 - m^2} + \xi \frac{L^{\mu \nu}}{q^2 - \xi m^2})$ leading to a 
$((n-1) A_0(m^2) + \xi A_0(\xi m^2))$ contribution, where $T^{\mu \nu}$ and $L^{\mu \nu}$ are the transverse 
and the longitudinal projectors and $n=2$ is the space-time dimension. For a true fermionic particle (massive leptons 
and quarks), we have $i \frac{\gamma.q + m}{q^2 - m^2} $ leading to a $m A_0(m^2)$ contribution.

To get the final results, one just has to sum up all the possible diagrams, taking into account the $-1$ for the fermionic 
loops (including Faddeev-Popov ghosts), the symmetry factor $s_i$ of the diagram $i$ and possibly the color factor 
for the quarks that we forget for conciseness. As a matter of fact, for the Higgs particle $h_0$ one gets:
$$ T_{h_0} = \Sigma_{i=1}^{9} c_i s_i t_i - \Sigma_{fermions} c_{10} s_{10} t_{10} - \Sigma_{i=11}^{12} c_i s_i t_i$$
where the couplings $c_i$, the symmetry factors $s_i$ and the propagator loops $t_i$ are given in the appendix. 
This formula leads, in a 2 dimensional space time, to a quite large formula depending on many parameters of the model 
as well as the mixing angles.

Similarly, for the other CP-even Higgs particle  $H^0$, one only has to change  $c_i, t_i$ into $C_i,T_i$
as indicated in the appendix where these coefficients are listed.

To summarise, the Veltman condition implies that the quadratic divergencies of the two possible tadpoles
$T_{h^0}$ and $T_{H^0}$ of the $h^0$ and $H^0$ CP-even neutral scalar fields vanish. Both results are not very tractable,
 so we do not write them. The linear combination of the fermionic coupling 
constants $s_{\alpha} c_{f \bar f} + c_{\alpha} C_{f \bar f}$ is zero; further it turns out that the combination 
$s_{\alpha} T_{h^0} + c_{\alpha} T_{H^0} $ induces simplification and one ends up with the short and nice expression:
\begin{equation}
T_t =  4 \frac{m_W^2}{v_{sm}^2} (\frac{1}{c_w^2} +1 ) + ( 2 \lambda_1 + 8 \lambda_2 +6 \lambda_3+ \lambda_4)
\end{equation}
where $c_w=\cos\theta_{weinberg}$ and $v_{sm}^2 = v_d^2 + 4 v_t^2 $ is the square of the Standard Model vacuum expectation value 
($\approx 246$ GeV).

\noindent
Of course, it is very tempting to calculate the orthogonal combination, 
$c_{\alpha} T_{h^0} - s_{\alpha} T_{H^0} $, which also leads to a simple result, where the quarks contributions 
must be multiplied by the color factor:
\begin{equation}
T_d =  - 2 Tr(I_n) \Sigma_f \frac{m_f^2}{v_d^2} + 3 (\lambda + 2 \lambda_1 + \lambda_4) 
+ 2 \frac{m_W^2}{v_{sm}^2} ( \frac{1}{c_w^2} +2 )
\end{equation}\\
Here, we do comment both results. The two linear combinations $s_{\alpha} h^0 + c_{\alpha} H^0 $ and $c_{\alpha} h^0 - s_{\alpha} H^0$ 
reproduce the real neutral components of the triplet and the doublet of the HTM, after their VEV shifts. 
So $T_t$ and $T_d$ represent the quadratic divergencies of the fields $\xi^0$ and $h$, as can be shown from Eq.  
(2.26) of \cite{aa11}. Hence, its is straightforward to understand that all mixing angles disappear in $T_t$ and $T_d$.

The absence of the $\lambda$ parameter in $T_t$ is obvious since $\lambda$ is the coupling constant of the 
pure doublet quartic interaction. In the same way, the absences of $\lambda_2$ and $\lambda_3$ are also 
natural in $T_d$ since these two couplings only concern the triplet. 

\noindent

We are also faced up to two intriguing questions: why the fermionic part is missing in the first equation?  
why the $\mu$ parameter is missing in both equations ?

The answer to the first question, on one hand, we have already noticed that the cancellation originates 
from the combination $s_{\alpha} T_h + c_{\alpha} T_H $, that is on the form of the trilinear couplings Higgs-fermion-antifermion. 
On the other hand, the absence of fermionic part lies in the form of the Yukawa coupling
with the triplet $\Delta$, given by Eq. (2.5) in \cite{aa11}). Indeed, a close look shows that the fermionic doublet $L$ is 
no longer associated with $ L^{\dag}$ but with $L^T$, that forbids  any $\xi^0 f \bar f$ coupling, 
insuring the absence of fermionic contribution at one loop order.

About the second question, it is bizarre because $\mu$ is the strength of a trilinear coupling between 
doublets and triplet, so a priori able to give contribution, but the interaction is linear in the triplet and a 
{\it doublet-triplet-triplet} vertex is excluded, that explains the lack of $\mu$ in $T_d$. As to  $T_t$, 
the solution is again in the form of the interaction where we have the doublet $H$ and the transposed 
doublet $H^T$ but not $ H^{\dag} $: before breaking of the symmetries and the vev shifts, a coupling 
such as  ${\delta^0}^*-\phi^0-{\phi^0}^*$ is forbidden. Then after breaking and shifting the fields, we 
end up with two vertices, $\xi^0-h-h$ and $\xi^0-Z_1-Z_1$, with opposite values $\pm \frac{\mu}{2 \sqrt 2}$, 
where $h$ and $Z_1$ are the real and imaginary parts of the shifted field $\phi^0$. This feature suffices 
to cancel these two contributions in a 2 dimensional space-time, when the loop does not depend on the propagator 
mass as explained before. In some sense, the quadratic divergencies record the physics before breaking and 
shifting, and this is probably not a pure one-loop effect.\\

Finally, it is worth to notice that Veltman Condition in Standard Model  \cite{Peyranere1991} is recovered 
when, we remove the couplings $\lambda_1$ and $\lambda_4$ in the doublet formula $T_d$, so discarding 
any mixing between doublet and triplet.

\section{Analysis: Implications for the Physical Scalar masses}
In this section we will focus our analysis on the modified Veltman condition (mVC) given in $Eqs. (25, 26)$. 
Our aim is twofold: first we will show that one loop quadratic divergencies are softened and go to zero within HTM space parameter. 
The latter is consistent with theoretical constraints given in Eqs. ($12-24$), namely unitarity, BFB as well as absence of 
tachyon in the potential (inducing a constraint on $\mu$ parameter). It also complies with the observed Higgs at $125$ GeV and with
 LHC measurements for Higgs decay to several channels ($\gamma \gamma$, $W\,W$, $Z\,Z$, $\tau \tau$, and  $b\,\bar{b}$) \cite{aa12}. 
Here we only show results with the photon mode that has the best mass resolution.  The second aim of our analysis is to gain more
 insight on the masses of the heavy Higgs bosons and the effect of naturalness on their allowed range of variations. \\

\begin{figure}[ht]
\centering
\resizebox{86mm}{!}{\includegraphics{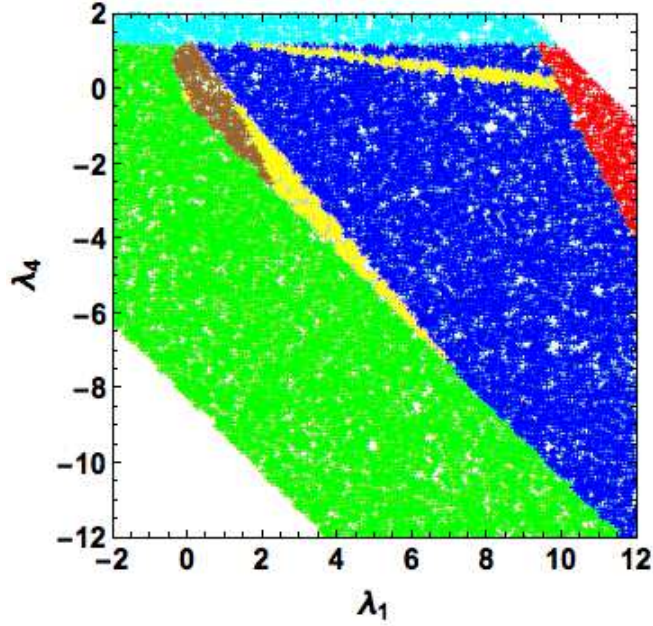}}
\caption{The allowed regions in ($\lambda_{1},\lambda_{4}$)
plans after imposing theoretical and experimental constraints. (\textcolor{cyan}{cyan}) : Excluded 
by $\mu$ constraints, (\textcolor{red}{red}) : Excluded by $\mu$+Unitarity constraints, 
(\textcolor{green}{green}) : Excluded by $\mu$+Unitarity+BFB constraints, (\textcolor{blue}{blue}) : 
Excluded by $\mu$+Unitarity+BFB+$R_{\gamma\gamma}$ constraints, (\textcolor{yellow}{yellow}) : 
Excluded by $\mu$+Unitarity+BFB $R_{\gamma\gamma}$\& $T_d=0$ $\land$ $T_t=0$ constraints. 
Only the brown area obeys  ALL constraints. Our inputs are $\lambda = 0.52$, 
$-2 \le \lambda_1 \le 12$, $\lambda_2 = -\frac{1}{6}$, $\lambda_3 = \frac{3}{8}$, 
$-12 \le \lambda_4 \le 2$, $v_t = 1$ GeV and $\mu = 1$ GeV.}
\label{fig:fig1}
\end{figure}

We plot in Fig.\ref{fig:fig1}, the scatter plot in the ($ \lambda_1,  \lambda_4$). The figure shows the excluded regions of space parameter 
by unitarity (red), by combined set of BFB and unitarity (green). If in addition, we impose consistency withe ATLAS and CMS combined 
data within $2 \sigma$ on the diphoton decay mode, (with a signal strength $R_{\gamma \gamma} = 1.15 \pm 0.25$ \cite{combined15}), the blue area is 
also discarded. Furthermore, when the Veltman condition for the doublet and triplet field enters into the game, we see a drastic 
reduction of the space parameter to a relatively small allowed region marked in brown.  \\

\begin{figure}[!h]
\centering
\resizebox{86mm}{!}{\includegraphics{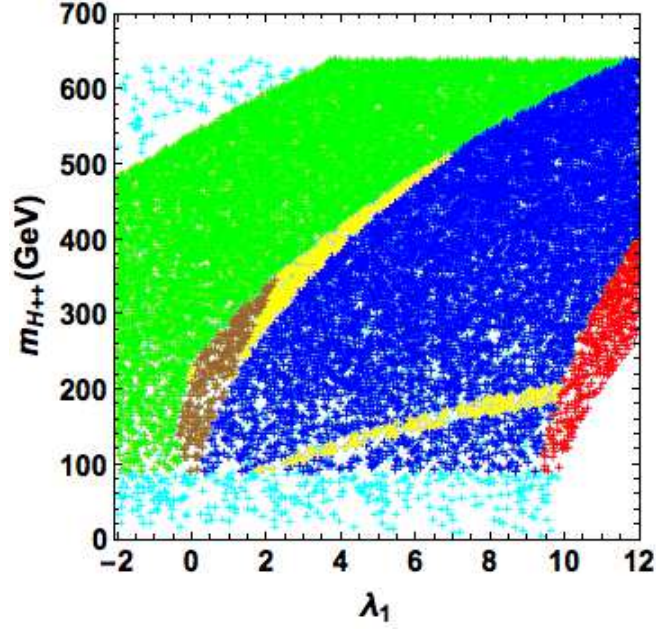}}
\caption{The allowed regions in ($\lambda_{1},m_{H^{\pm\pm}}$) 
plans after imposing theoretical and experimental 
constraints. Our inputs are $\lambda = 0.52$, $-2 \le \lambda_1 \le 12$, 
$\lambda_2 = -\frac{1}{6}$, $\lambda_3 = \frac{3}{8}$, $-12 \le \lambda_4 \le 2$, 
$v_t = 1$ GeV and $\mu = 1$ GeV.}
\label{fig:fig2}
\end{figure}

Fig.\ref{fig:fig2} illustrates the doubly charged Higgs mass $m_{H^{\pm\pm}}$ as a function of  $\lambda_1$ resulting from a scan over different values of $\lambda_4$ and for $\mu = v_t = 1$. We find the only relevant parameter region where all of the constraints are imposed, is the area marked in grey which encodes cancellation of quadratic divergencies. This area
 is delineated by $\lambda_1$ in the range[$-0.3, 2.3$] corresponding to $m_{H^{\pm \pm}}$ varying from $90$ to $351$ ~GeV. For these
 mass values,  $\lambda_4$ is limited to lie in a reduced interval between $-2.6$ and $1.18$. \\
 
\begin{figure}[!h]
\begin{tabular}{rr}
\hspace{-.8cm}\resizebox{86mm}{!}{\includegraphics{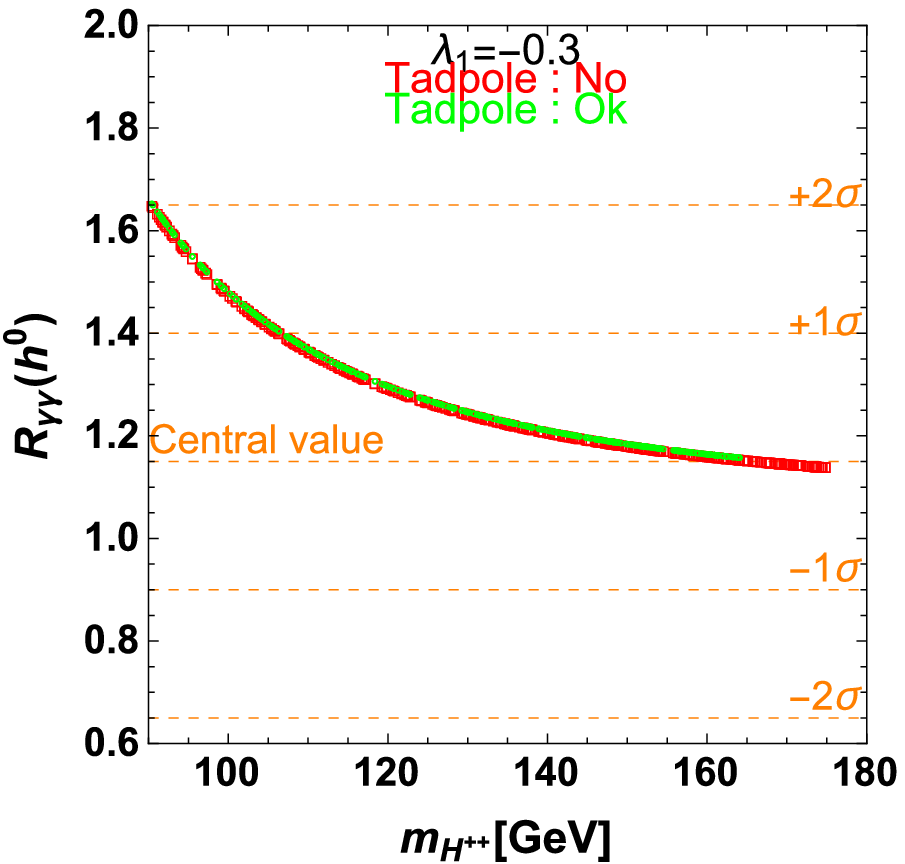}}&
\hspace{.1cm}\resizebox{85mm}{!}{\includegraphics{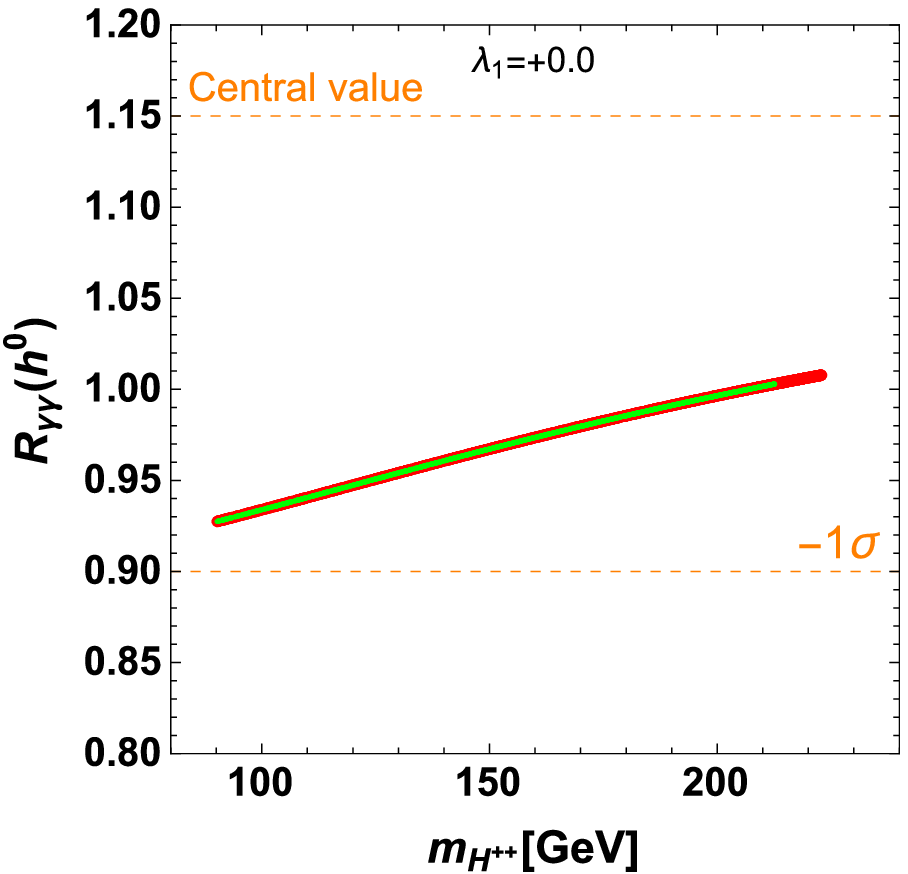}}\\
\hspace{-.8cm}\resizebox{86mm}{!}{\includegraphics{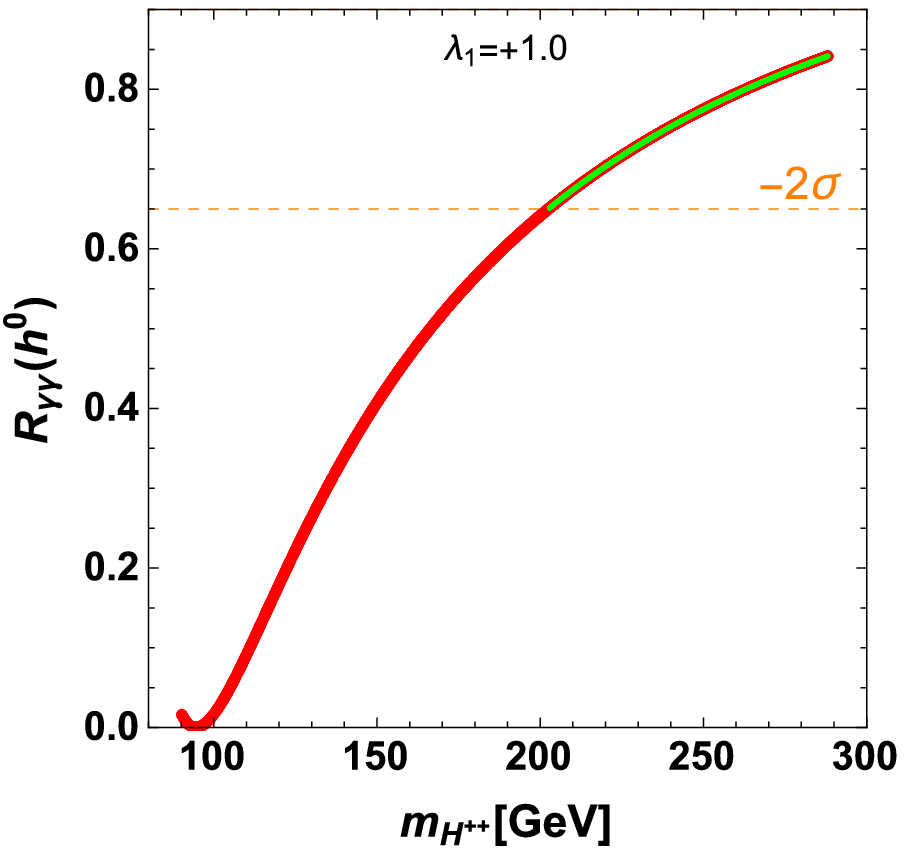}}&
\hspace{.1cm}\resizebox{85mm}{!}{\includegraphics{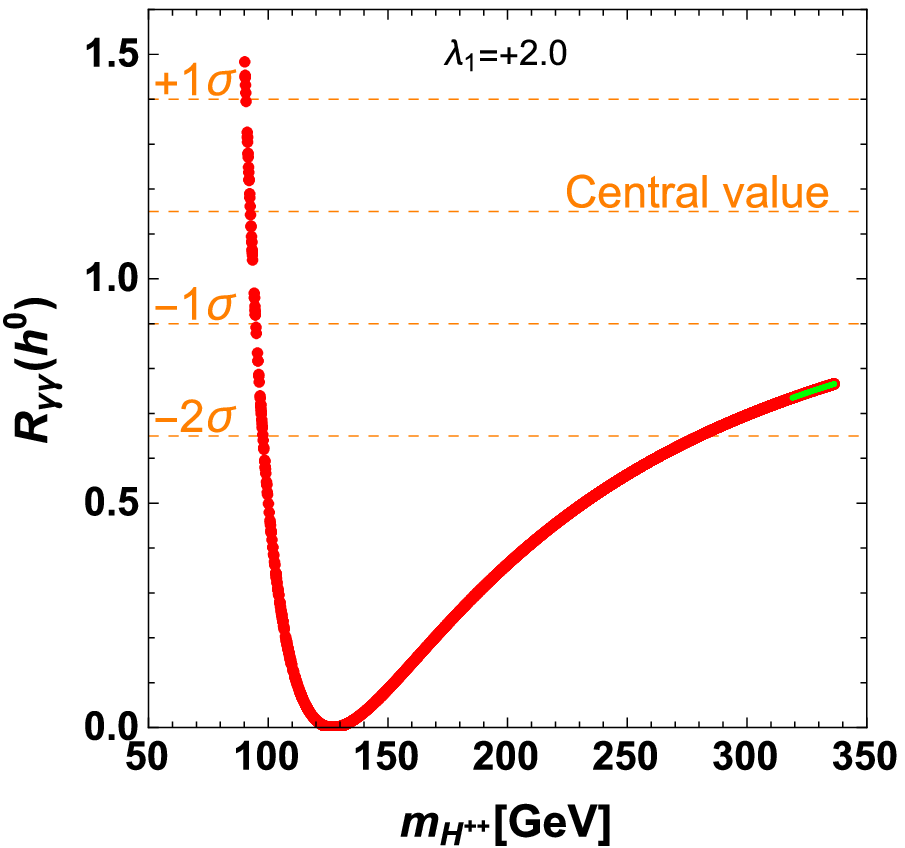}}
\end{tabular}
\caption{$R_{\gamma\gamma}(h^0)$ as a function of $m_{H^{\pm\pm}}$ 
for various values of $\lambda_1$ with and without Veltman conditions 
($T_d=0$ $\land$ $T_t=0$). We scan over the HTM parameters as : 
$\lambda = 0.52$, $\lambda_2 = -\frac{1}{6}$, $\lambda_3 = \frac{3}{8}$, 
$-10 \le \lambda_4 \le 2$, $v_t = 1$ GeV and $\mu = 1$ GeV.}
\label{fig:fig3}
\end{figure}

\begin{figure}[!h]
\hspace{-.8cm}\resizebox{86mm}{!}{\includegraphics{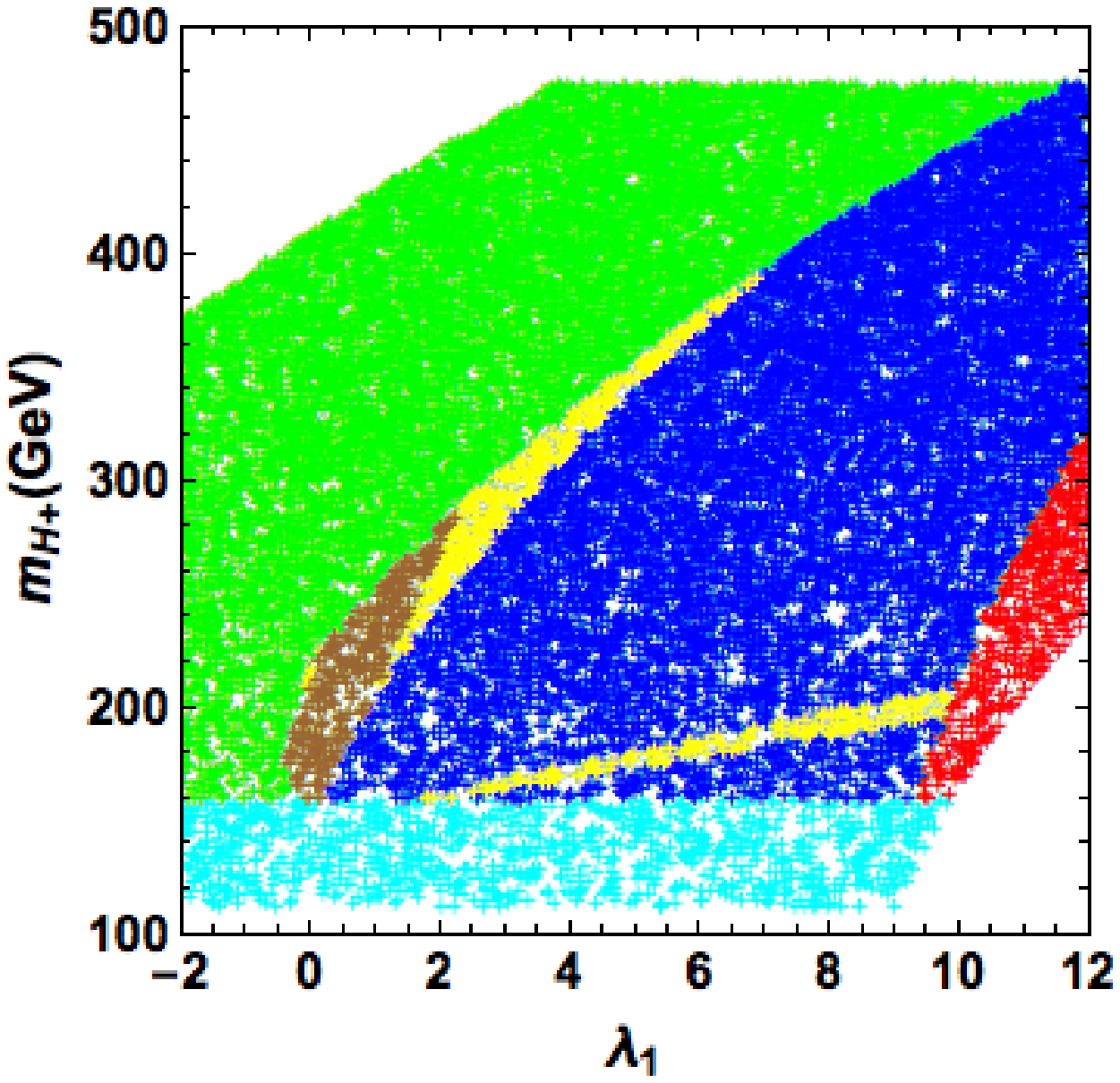}}\\
\hspace{-.8cm}\resizebox{84.5mm}{!}{\includegraphics{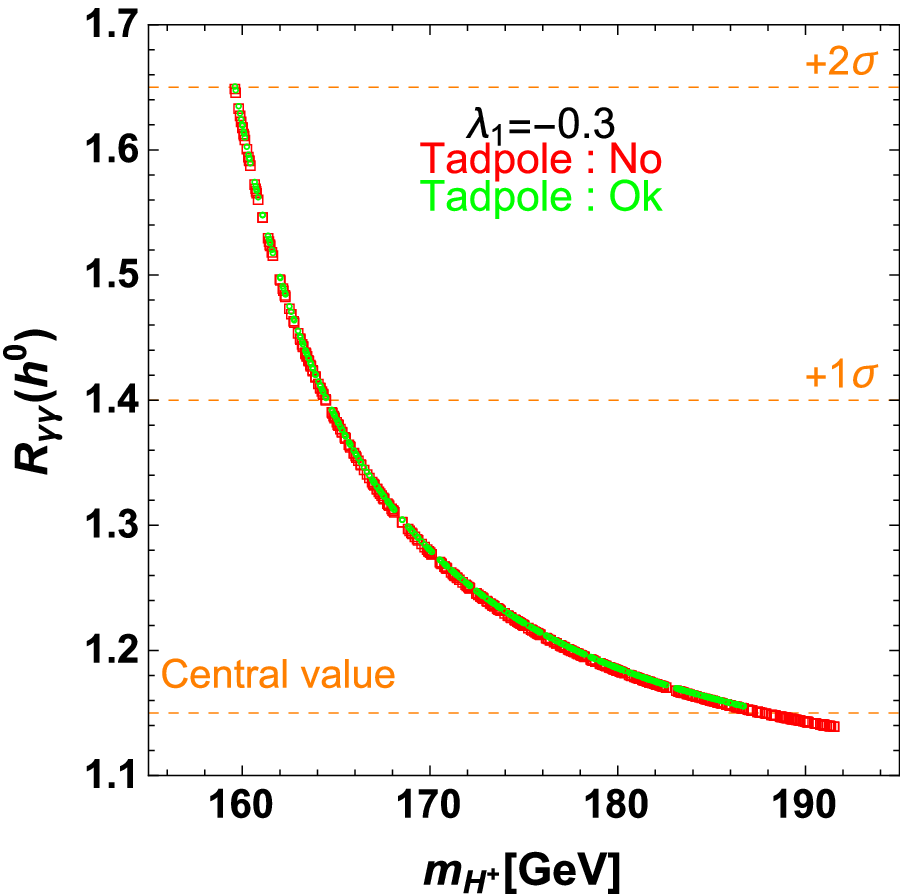}}
\hspace{-.8cm}\resizebox{86mm}{!}{\includegraphics{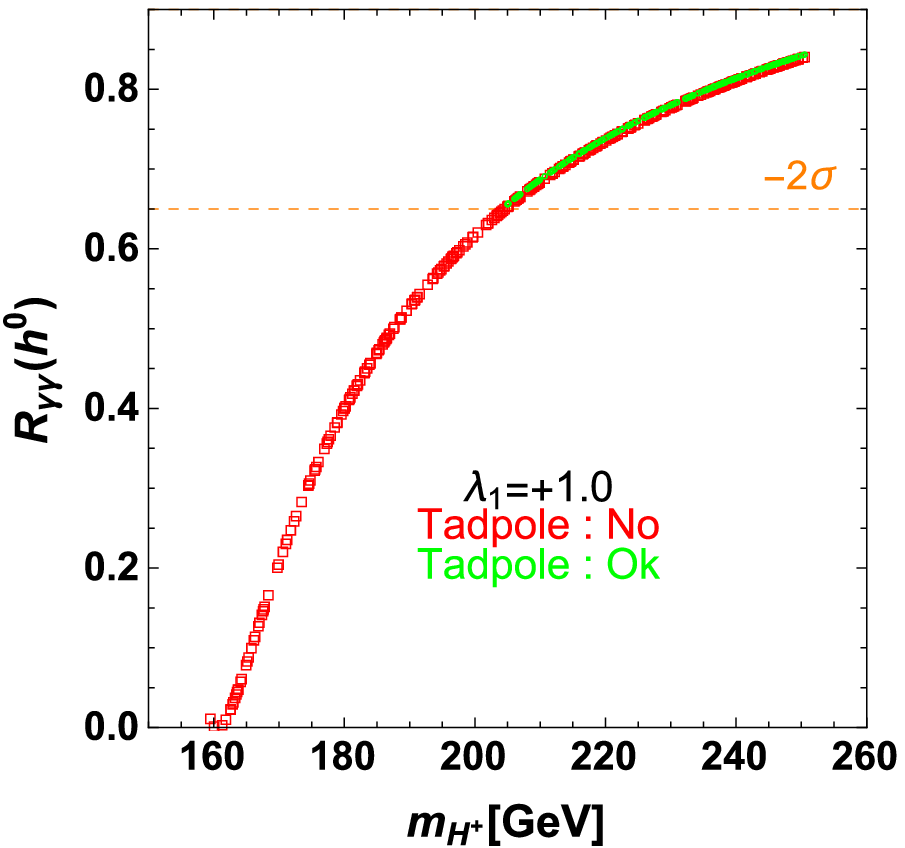}}
\caption{Upper slide : The allowed regions in ($\lambda_{1},m_{H^{\pm}}$) plans 
after imposing theoretical and experimental constraints. Lower 
slide : $R_{\gamma\gamma}(h^0)$ as a function of $m_{H^{\pm}}$ for $\lambda_1 = -0.3$ 
(left) and $\lambda_1 = +1.0$ (right)  with and without Veltman conditions 
($T_d=0$ $\land$ $T_t=0$). Our inputs are $\lambda = 0.52$,
$\lambda_2 = -\frac{1}{6}$, $\lambda_3 = \frac{3}{8}$, $-12 \le \lambda_4 \le 2$ and
$\mu = v_t = 1$ GeV.}
\label{fig:fig4}
\end{figure}

The remarkable feature of effects of the modified Veltman condition on the doubly charged Higgs masse is clearly indicated in Fig.\ref{fig:fig3} showing the $R_{\gamma\gamma}$ ratio (signal strength) as a function of the doubly charged Higgs mass $m_{H^{\pm\pm}}$ for different values of $\lambda_1$. First we consider the case when when mVC are absent (red plot). We see that, small values of $\lambda_1$ less 
than $1$ favour low $m_{H^{\pm\pm}}$ varying between $90$ and $230$~GeV, while for larger $\lambda_1$,  its range  of variation is significantly enlarged with an upper limit about $515$~GeV. When mVC is turned on (green plot), we show that the $m_{H^{\pm\pm}}$ intervals get smaller and smaller for $\lambda_1$ getting larger. We also note that unlike the lower limit which is unaffected by mVC, 
the upper bound is very sensitive to mVC and experiment a drastic reduction to $351$~GeV.  Similar analysis is seen in Fig.\ref{fig:fig4} for the simply charged Higgs mass. Here, the same behaviour is reproduced: the lower mass, $160$~GeV, is almost insensitive to the constraints including mVC, while the upper limit shows a substantial reduction from $392$~GeV to $288$~GeV when the conditions mVC are considered. These overall resulting ranges of  $m_{H^{\pm}}$ and  $m_{H^{\pm\pm}}$ are compatible with the LHC exclusion limits on the charged Higgs bosons ($H^{\pm}, H^{\pm\pm}$) . \\
 
 \section{Conclusion}
 We have studied the naturalness problem In the type II seesaw model. We have shown 
that the Veltman condition is modified by virtue of the additional scalar charged states and that quadratic divergencies at one 
loop can be driven to zero within at the allowed parameter space of the model. Furthermore, when a set of constraints, including unitarity, consistent boundedness from below, is combined with requirement of compatibility with the diphoton Higgs decay data of LHC for the observed Higgs at $125$~GeV, the resulting parameter region is severely constrained. If in addition, the modified Veltman condition is also imposed, the effects on the masses of heavy Higgs bosons $H^0$, $A^0$, $H^\pm$ and 
$H^{\pm\pm}$ are analysed. The analysis has shown a drastic reduction of the mass spectrum of $m_{H^{\pm}}$ and $m_{H^{\pm\pm}}$ with an upper bounds at $288$ and $351$~GeV respectively. These limits are still consistent with LHC measurements . Besides, we have found an almost mass degeneracy of CP odd Higgs and heavy CP even Higgs, with $m_{H^{0}} \simeq  m_{A^{0}} \simeq 207~GeV$. \footnote{An diphoton excess around $750$~GeV has been reported  by ATLAS and CMS with a local significance of $3.6 \sigma $ and $2.6 \sigma $ while the global significances are $2 \sigma$ and $1.2 \sigma$ respectively \cite{atlas_ex, cms_ex}.  Although several interpretations of this intriguing signal in terms of Higgs-like resonances have been proposed, this excess could be just another statistical fluctuation which will be washed away with more data.}

\section*{Acknowledgements}
M.~Chabab acknowledges support from the grant ${\rm H2020}-{\rm MSCA}-{\rm RISE}/2014-645722$. This work is supported in part by the Moroccan Ministry of Higher Education and Scientific Research under contract ${\rm n}^o\,{\rm PPR}/2015/6$. M.~C. ~Peyranere would like to thank Gilbert Moultaka for friendly and helpful discussions.

\newpage
\section*{Appendix : Feynman Rules}
\label{sec:appendix}

In this appendix, we give the couplings used to calculate the tadpoles of the two neutral CP-even Higgs $h^0$ and $H^0$.
Since we are interested in the one-loop contributions, only three-leg couplings are useful. Further, within this  
restricted class, we look for vertices such as $h^0 F_i \bar {F_i}$ or  $H^0 F_i \bar {F_i}$, where $F_i$ stands for any quantum field
of our model: scalar and vectorial bosons, fermions, Goldstone fields $G_i$ and Faddeev-Popov ghost fields $ \eta_i$.
To be precise, we have used the well-known linear $R_{\xi}$ gauge where the gauge fixing Lagrangians are
$\frac{-1}{2\xi_Z} (\partial_{\mu} Z^{\mu} -\xi_Z m_Z G_0)^2$ for the neutral sector and 
$\frac{-1}{2\xi_W}(\partial_{\mu} W_{\pm}^{\mu} -\xi_W m_W G_{\pm})^2$ for the charged sector. \\

We note $c_{F_i \bar {F_i}}$ ($C_{F_i \bar {F_i} }$) the couplings to the Higgs $h^0$ ($H^0$). Since the field $F_i$ fixes the propagator,
 we also give the values $t_i$ ($T_i$) of the loop due to the propagator of the $F_i$ particle which gain a factor   
 $2$ in case of charged fields, and the symmetry factor $s_i$.
\begin{eqnarray}
&& c_1 \equiv c_{h_0h_0} = \frac{-3 i}{2} (c_\alpha^3 \lambda v_d + 2 c_\alpha (\lambda_1 + \lambda_4) s_\alpha^2 v_d + 
      4 (\lambda_2 + \lambda_3) s_\alpha^3 v_t + 2 c_\alpha^2 s_\alpha (-\sqrt {\mu} + (\lambda_1 + \lambda_4) v_t)), \nonumber\\
&& C_1 \equiv C_{H_0 H_0} = \frac{3 i}{2} (2 c_\alpha^2 (\lambda_1 + \lambda_4) s_\alpha v_d + \lambda s_\alpha^3 v_d - 
      4 c_\alpha^3 (\lambda_2 + \lambda_3) v_t + 2 c_\alpha s_\alpha^2 (\sqrt 2 \mu - (\lambda_1 + \lambda_4) v_t)), \nonumber\\
&& t_1 = i A_0(m_{h_0}^2),\nonumber\\
&& T_1 = i A_0(m_{H_0}^2), \nonumber\\
&& s_1= \frac{1}{2},
\end{eqnarray} 
\begin{eqnarray}
&& c_2 \equiv c_{G_0 G_0} = -\frac{ i}{2} (-4 \sqrt{2} \mu c_\alpha c_\beta s_\beta + 2 s_\beta^2 (c_\alpha (\lambda_1 + \lambda_4) v_d + 2 (\lambda_2 + \lambda_3) s_\alpha v_t) + c_\beta^2 (2 \sqrt{2} \mu s_\alpha + c_\alpha \lambda v_d \nonumber\\
&&\hspace{2.cm} + 2 (\lambda_1 + \lambda_4) s_\alpha v_t)),\nonumber\\
&& C_2 \equiv C_{G_0 G_0} = -\frac{i}{2} (s_\alpha (4 \sqrt 2 c_{\beta} \mu s_\beta - c_{\beta}^2 \lambda v_d - 2 (\lambda_1 + \lambda_4) s_\beta^2 v_d) + 2 c_\alpha (2 (\lambda_2 + \lambda_3) s_\beta^2 v_t \nonumber\\
&&\hspace{2.cm}  + c_{\beta}^2 (\sqrt 2 \mu + (\lambda_1 + \lambda_4) v_t))),\nonumber\\
&& t_2 = T_2 = i A_0(\xi_Z m_Z^2),\nonumber\\
&& s_2= \frac{1}{2},
\end{eqnarray}
\begin{eqnarray}
&& c_3 \equiv c_{G_+ G_-} = -\frac{i}{2}(s_{\beta'}^2 (c_\alpha (2 \lambda_1 + \lambda_4) v_d + 
4 (\lambda_2 + \lambda_3) s_\alpha v_t) + c_{\beta'}^2 (c_\alpha \lambda v_d + 2 \lambda_1 s_\alpha v_t) - c_{\beta'} s_{\beta'} \nonumber\\
&&\hspace{2.cm}  (4 c_\alpha \mu - \sqrt{2} \lambda_4 (s_\alpha v_d + c_\alpha v_t))) , \nonumber\\
&& C_3 \equiv C_{G_+ G_-} = -\frac{i}{2}(s_{\beta'}^2 ((2 \lambda_1 + \lambda_4 ) s_\alpha v_d - 4 c_\alpha (\lambda_2 + \lambda_3) v_t) + c_{\beta'}^2 (\lambda s_\alpha v_d - 2 c_\alpha \lambda_1 v_t) - c_{\beta'} s_{\beta'} \nonumber\\
&& \hspace{2.cm} (4 \mu s_\alpha + \sqrt 2 \lambda_4) (c_\alpha v_d - s_\alpha v_t))),\nonumber\\
&& t_3  = T_3 = 2 \times i A_0(\xi_W m_W^2),\nonumber\\
&& s_3= \frac{1}{2},\nonumber\\
\end{eqnarray}
\begin{eqnarray}
&& c_4 \equiv c_{H_0H_0} = -\frac{i}{2} (2 c_\alpha^3 (\lambda_1 + \lambda_4) v_d + 
c_\alpha (3 \lambda - 4 (\lambda_1 + \lambda_4)) s_\alpha^2 v_d +  
2 s_\alpha^3 (-\sqrt{2} \mu + (\lambda_1 + \lambda_4) v_t)  \nonumber\\
&&\hspace{2.cm}  + 4\,c_\alpha^2\,s_\alpha (\sqrt{2} \mu - (\lambda_1 - 3 (\lambda_2 + \lambda_3) + \lambda_4) v_t)) \nonumber\\
&& C_4 \equiv C_{h_0 h_0} = \frac{ i}{2}(c_\alpha^2 (3 \lambda - 4 (\lambda_1 + \lambda_4)) s_\alpha v_d + 
  2 (\lambda_1 + \lambda_4) s_\alpha^3 v_d + 2 c_\alpha^3 (\sqrt 2 \mu - (\lambda_1 + \lambda_4) v_t) \nonumber\\
&&\hspace{2.cm} - 4\,c_\alpha\,s_\alpha^2 (\sqrt 2 \mu - (\lambda_1 - 3 (\lambda_2 + \lambda_3) + \lambda_4) v_t)), \nonumber\\
&& t_4= i A_0(m_{H_0}^2), \nonumber\\
&& T_4= i A_0(m_{h_0}^2), \nonumber\\
&& s_4= \frac{1}{2},
\end{eqnarray}
\begin{eqnarray}
&& c_5 \equiv c_{A_0A_0}  = -\frac{i}{2} (c_\alpha (4 \sqrt{2} c_\beta \mu s_\beta + 2 c_\beta^2 (\lambda_1 + \lambda_4) v_d + \lambda s_\beta^2 v_d) + 2 s_\alpha (2 c_\beta^2 (\lambda_2 + \lambda_3) v_t + s_\beta^2 (\sqrt{2} \mu \nonumber\\
&&\hspace{1.5cm} + (\lambda_1 + \lambda_4) v_t))),\nonumber\\
&& C_5 \equiv C_{A_0A_0}  = \frac{ i}{2}(4 \sqrt 2 c_{\beta} \mu s_\alpha s_\beta + 2 c_{\beta}^2 ((\lambda_1 + \lambda_4) s_\alpha v_d - 2 c_\alpha (\lambda_2 + \lambda_3) v_t) + s_\beta^2 (\lambda s_\alpha v_d - 2 c_\alpha (\sqrt 2 \mu\nonumber\\
&&\hspace{1.5cm}   + (\lambda_1 + \lambda_4) v_t))),\nonumber\\
&& t_5 = T_5 = i A_0(m_{A_0}^2),\nonumber\\
&& s_5= \frac{1}{2},
\end{eqnarray}
\begin{eqnarray}
&& c_6 \equiv c_{H_+ H_-} =  -\frac{i}{2}(c_{\beta'}^2 (c_\alpha (2 \lambda_1 + \lambda_4) v_d + 4 (\lambda_2 + \lambda_3) s_\alpha v_t) + s_{\beta'}^2 (c_\alpha \lambda v_d + 2 \lambda_1 s_\alpha v_t) + c_{\beta'} s_{\beta'}(4 c_\alpha \mu\nonumber\\
&&\hspace{2.0cm} - \sqrt{2} \lambda_4 (s_\alpha v_d + c_\alpha v_t))),\nonumber\\
&& C_6 \equiv C_{H_+ H_-} = \frac{i}{2}(c_{\beta'}^2 ((2 \lambda_1 + \lambda_4 ) s_\alpha v_d - 
4 c_\alpha (\lambda_2 + \lambda_3) v_t) + s_{\beta'}^2 (\lambda s_\alpha v_d - 2 c_\alpha \lambda_1 v_t) + c_{\beta'} s_{\beta'} (4 \mu s_\alpha \nonumber\\
&&\hspace{2.0cm} + \sqrt 2 \lambda_4) (c_\alpha v_d - s_\alpha v_t))),\nonumber\\
&& t_6 = T_6 = 2 \times i A_0(m_{H_{\pm}}^2),\nonumber\\
&& s_6= \frac{1}{2},
 \end{eqnarray} 
\begin{eqnarray}   
&& c_7 \equiv c_{H_{++} H_{--}} = -i (c_\alpha \lambda_1 v_d + 2 \lambda_2 s_\alpha v_t),\nonumber\\
&& C_7 \equiv C_{H_{++} H_{--}} = i (\lambda_1  s_\alpha v_d - 2 c_\alpha \lambda_2 v_t),\nonumber\\
&& t_7 = T_7 = 2 \times i A_0(m_{H_{\pm\pm}}^2),\nonumber\\
&& s_7= \frac{1}{2},\nonumber\\
 \end{eqnarray}
\begin{eqnarray}
&& c_8 \equiv c_{ZZ} = (i e m_W (c_\alpha c_{\beta'} + 2 \sqrt{2} s_\alpha s_{\beta'}))/(c_w^2 s_w),\nonumber\\
&& C_8 \equiv C_{ZZ} = (-i e m_W (c_{\beta'} s_\alpha - 2 \sqrt 2 c_\alpha s_{\beta'}))/(c_w^2 s_w),\nonumber\\
&& t_8 = T_8 = -i((n-1) A_0(m_Z^2) + \xi_Z A_0(\xi_Z m_Z^2),\nonumber\\
&& s_8= \frac{1}{2},
\end{eqnarray}
\begin{eqnarray}
&& c_9 \equiv c_{W_+ W_-} = i e m_W (c_\alpha c_{\beta'} + \sqrt{2} s_\alpha s_{\beta'})/s_w,\nonumber\\
&& C_9 \equiv C_{W_+ W_-} =-i e m_W (c_{\beta'} s_\alpha -\sqrt 2 c_\alpha s_{\beta'})/s_w,\nonumber\\
&& t_9 = T_9 = 2 \times (-i((n-1) A_0(m_W^2) + \xi_W A_0(\xi_W m_W^2)),\nonumber\\
&& s_9= \frac{1}{2},
\end{eqnarray}
\begin{eqnarray}
&& c_{10} \equiv c_{f \bar f}  = \frac{-i}{2} e (c_\alpha/c_{\beta'}) m_f/(m_W s_w),\nonumber\\
&& C_{10} \equiv C_{f \bar f}  = \frac{i}{2}  e (s_\alpha/c_{\beta'}) m_f/ m_W s_w),\nonumber\\
&& t_{10} = T_{10} = i m_f A_0(m_f^2) Tr(I_n),\nonumber\\
&& s_{10}= 1,
\end{eqnarray}
\begin{eqnarray}
&& c_{11} \equiv c_{{\eta_Z} \bar{\eta_Z}}=  \frac{-i}{2} e m_W (c_\alpha c_{\beta'} + 2 \sqrt{2} s_\alpha s_{\beta'}) \xi_Z)/(c_w^2 s_w),\nonumber\\
&& C_{11} \equiv C_{{\eta_Z} \bar{\eta_Z}}=  \frac{ i}{2} e m_W (c_{\beta'} s_\alpha - 2 \sqrt{2} c_\alpha s_{\beta'}) \xi_Z/(c_w^2 s_w),\nonumber\\
&& t_{11} = T_{11} = i A_0(\xi_Z m_Z^2),\nonumber\\
&& s_{11}= 1,
\end{eqnarray}
\begin{eqnarray}
&& c_{12} \equiv c_{\eta_{\pm} \bar{\eta_{\pm}}} = \frac{-i}{2} e m_W (c_\alpha c_{\beta'} + \sqrt{2} s_\alpha s_{\beta'}) \xi_W/s_w,\nonumber\\
&& C_{12} \equiv C_{\eta_{\pm} \bar{\eta_{\pm}}} = \frac{ i}{2} e m_W (c_{\beta'} s_\alpha - \sqrt{2} c_\alpha s_{\beta'})\xi_W/s_w,\nonumber\\ 
&& t_{12} = T_{12} = 2 \times i A_0(\xi_W m_W^2),\nonumber\\
&& s_{12}= 1,
\end{eqnarray}

\begin{table}[!h]
\begin{center}
\renewcommand{\arraystretch}{1.5}
\begin{tabular}{|p{1.10cm}|p{3.3cm}|p{3.cm}|p{3.cm}|p{3.cm}|}
\hline
\hline
$m_{\Phi}$ &  {\bf Unitarity} & {\bf Unitarity + \vspace{-0.1cm}\newline\vspace{-0.1cm}\hspace{-0.1cm} BFB}  &  {\bf Unitarity +\vspace{-0.1cm}\newline\vspace{-0.1cm} \hspace{-0.1cm}BFB +\vspace{-0.1cm}\newline\vspace{-0.1cm} \hspace{-0.1cm}$R_{\gamma\gamma}$} &  {\bf Unitarity +\vspace{-0.1cm}\newline\vspace{-0.1cm} \hspace{-0.1cm}BFB +\vspace{-0.1cm}\newline\vspace{-0.1cm} \hspace{-0.1cm}$R_{\gamma\gamma}$ +\vspace{-0.1cm}\newline\vspace{-0.1cm} \hspace{-0.1cm}mVC} \\
\hline
\hline
$H^0$ &  $[206.8-207.3]$ GeV &  $[206.8-207]$ GeV  &  $[206.8-207]$ GeV  &  $206.8$ GeV  \\ 
\hline
\hline
$A^0$ &  $206.8$ GeV &  $206.8$ GeV &  $206.8$ GeV &  $206.8$ GeV  \\ 
\hline
\hline
$H^\pm$ & $[160-474]$ GeV  &  $[160-474]$ GeV  &  $[160-392]$ GeV  &  $[161-288]$ GeV  \\ 
\hline
\hline
$H^{\pm\pm}$ &  $[90-637]$ GeV  &  $[90-637]$ GeV   &  $[90-513]$ GeV   &  $[90-351]$ GeV   \\ 
\hline
\end{tabular}
\end{center}
\caption{Higgs bosons masses allowed  intervals in the Higgs triplet model resulting from various constraints, including the modified Veltman conditions}
\label{table}
\end{table}


\newpage

\end{document}